\documentclass[preprint,5p,times,twocolumn]{elsarticle}
\journal{Nuclear Instrumentation Method A}
\usepackage{amssymb}
\usepackage{siunitx}
\usepackage{lineno}
\usepackage{xspace}
\usepackage{floatrow}
\usepackage{hyperref}
\usepackage{natbib}
\usepackage{subfig}
\usepackage{upgreek}
\usepackage{caption}
\usepackage{caption3}

\def\vbias {V$_{\mathrm{bias}}$\xspace}
\def\vone    {V$_1$\xspace}
\def\vtwo    {V$_2$\xspace}
\def\dv    {$\Delta$V\xspace}
\def\bary  {$<\mathrm{Z}>_{\mathrm{E}}$\xspace}
\def\drift     {\raisebox{.5pt}{\textcircled{\raisebox{-.9pt} {1}}}\xspace}
\def\induction {\raisebox{.5pt}{\textcircled{\raisebox{-.9pt} {2}}}\xspace}
\def\electrodeone{electrode \textbf{1}\xspace}
\def\electrodetwo{electrode \textbf{2}\xspace}
\def\neq {n$\mathrm{_{eq}}$/cm$^{2}$\xspace}

\begin{document}
\begin{frontmatter}
\title{The Silicon Electron Multiplier Sensor\tnoteref{t1}}



\author[a,b]{Marius M\ae hlum Halvorsen\corref{cor1}}
\author[a]{Victor Coco}
\author[a]{Evangelos Leonidas Gkougkousis}
\author[a]{Paula Collins}
\author[c]{Olivier Girard}

\address[a]{{CERN EP-LBD}, {Esplanade des Particules 1}, {Meyrin}, {1211}, {Switzerland}}
\address[b]{{University of Oslo, Department of Physics}, {Sem S\ae lands vei 24}, {Oslo}, {0315}, {Norway}}
\address[c]{{CEA-Leti}, {17 avenue de Martyrs}, {Grenoble}, {38054}, {France}}


\cortext[cor1]{Corresponding author}

\begin{abstract}
The Silicon Electron Multiplier (SiEM) is a novel sensor concept for minimum ionizing particle (MIP) detection which uses internal gain and fine pitch to achieve excellent temporal and spatial resolution.  In contrast to sensors where the gain region is induced by doping (LGADs, APDs), amplification in the SiEM is achieved by applying an electric potential difference in a composite electrode structure embedded within the silicon bulk using MEMS fabrication techniques. Since no gain-layer deactivation is expected with radiation damage, such a structure is expected to withstand fluences of up to 10$^{16}$\neq. Various geometries and biasing  configurations are studied, within the boundaries imposed by the fabrication process being considered. The effective gain, the field in the sensor, the leakage current and breakdown conditions are studied for cell sizes in the range of 6 – \SI{15}{\micro\meter}. Simulations show that gains in excess of 10 can be achieved, and studies of the time structure of the induced signals from a charge cloud deposited in the middle of the sensor show that time resolutions similar to other sensors with internal gain can be expected.   Plans for the manufacture of a proof-of-concept sensor and for its subsequent characterisation are discussed.
\end{abstract}
\end{frontmatter}

\section{Introduction}
\label{introduction}
Charged particle detection in the innermost region of high energy physics collider experiments combines the challenges of high fluence and high occupancy with the need for measurements with excellent spatial and time resolution. While demand for spatial resolutions below \SI{10}{\micro \meter} is now commonplace, an ever increasing number of applications now require in addition a time resolution below 50~ps~\cite{CERN-LHCC-2020-007,CMS:2667167,Aaij:2244311,na62,tauLV}.
Existing solid state detector technologies suffer various limitations. While thin planar technology allows fluences of several times 10$^{16}$cm$^{2}$s$^{-1}$ to be reached, and can in principle reach time resolution as low as 30~ps~\cite{Riegler_2017}, the limited signal amplitude makes this difficult to exploit. High fluences can also be reached with 3D technologies, and thanks to the decoupling between the drift direction and charge deposition direction, the Landau fluctuation contribution to the time resolution becomes sub-leading. Hence, time resolution below the 20-30~ps range have been measured~\cite{timespot2020,cnm3d}. Nonetheless, 3D technologies suffer high capacitance, fill factor reduction and it will be difficult to reach pitches lower than \SI{50}{\micro\meter} with present technologies. 
Another approach consists of amplifying the signal within the sensors, such as done in the  LGADs~\cite{PELLEGRINI201412} family. Gains of around 10 allow time resolutions in the range 20 – 30$\,$ps to be reached. Recent variations of the LGAD design address the limitation in segmentation (iLGAD~\cite{PELLEGRINI201624}, trenched LGAD, AC-LGADs, \ldots). However LGADs suffer gain loss for fluences above 2 10$^{15}$\neq~\cite{Mazza_2020} due to acceptor removal in the p+-doped gain layer.

We propose here a novel sensor design, the silicon electron multiplier (SiEM), which provides intrinsic gain together with a pixel pitch better than \SI{10}{\micro\meter}, without reliance on shallow doping.  The SiEM is expected to be more radaition hard than LGADs while providing similar time resolution. 

\section{Sensor description}
\label{sec:description}
\begin{figure}[h!]
    \centering
    \includegraphics[width=0.96\textwidth]{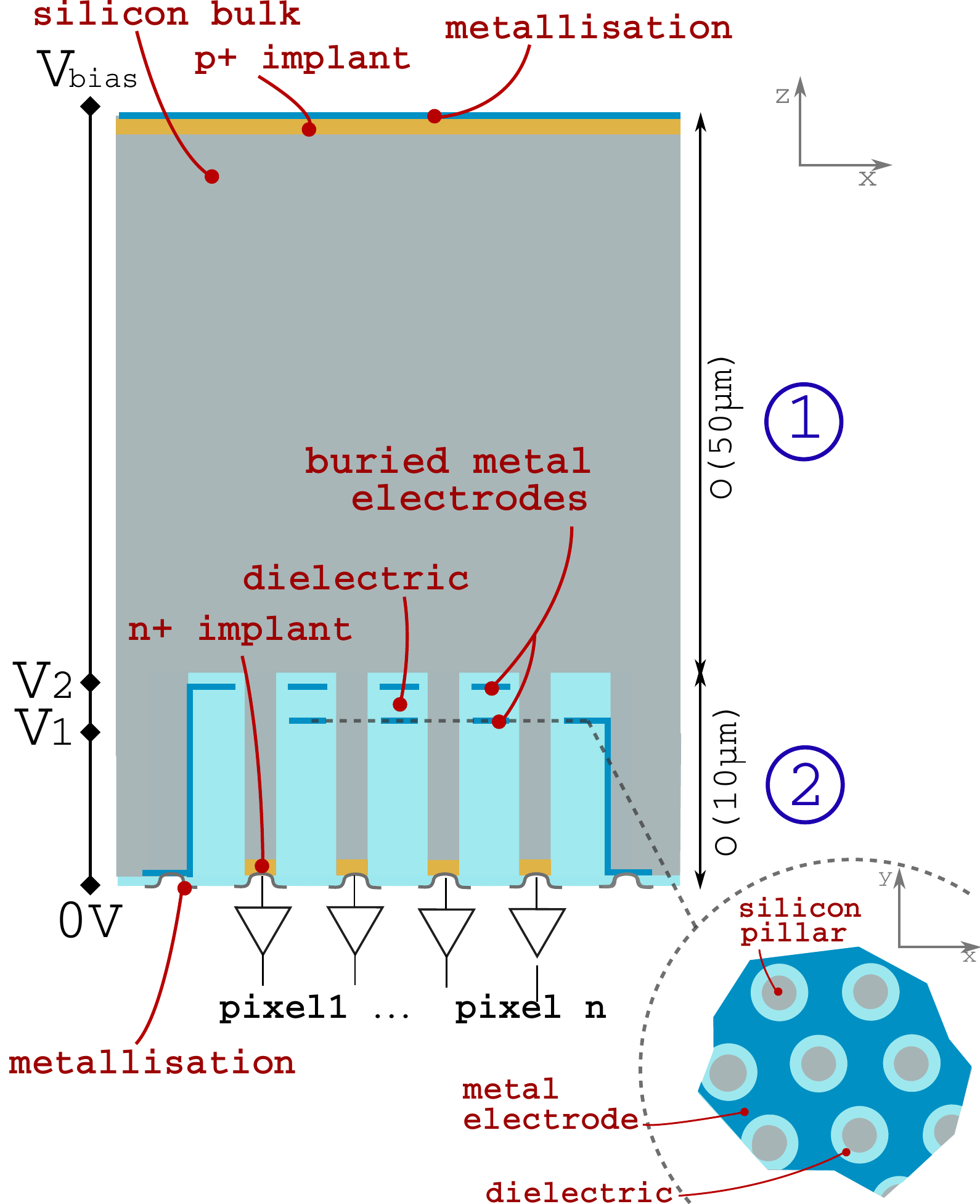}
    \caption{Schematic description of the SiEM detector. It consists of an O(\SI{50}{\micro \meter}) thick depleted silicon region \drift adjacent to a region \induction made of silicon pillars. In between the pillars two metallic electrodes, separated by a dielectric are buried.
    The electrodes produce a high electric field region in \induction. As a particle passes through \drift, primary electrons are produced by ionisation and drift toward \induction where they are multiplied and induce a signal while drifting towards the readout electrode.}
    \label{fig:geometry}
\end{figure}

\subsection{Detection principle}
\label{sec:description:det_principle}

As shown in Figure~\ref{fig:geometry} the SiEM can be delineated into two regions. In a conversion and drift layer, primary charge carriers are created by the passage of minimum ionising particles (MIPs) through the sensor which then drift under the influence of an electric field towards an amplification and induction layer. In this layer a region of high electric field is generated in the material to achieve primary charge multiplication by impact ionisation. The high electric field is achieved by applying electric potentials to a composite conductive structure buried in the bulk of the material. The drift of the secondary charges in this region induces a signal on the readout electrode.
The exact geometry of the amplification and induction regions depends on the  bulk material as well as the fabrication process that can be achieved. For the design proposed in this paper, as illustrated in Figure~\ref{fig:geometry}, the bulk of the sensor consists of a reversed biased n-in-p silicon junction. The readout side is etched, leaving cylindrical Si-pillars which are terminated with the n+ implant and metal contact to the readout electrodes. A composite structure consisting of two conductive grid electrodes separated by a dielectric is deposited in the etched region. A difference of potential is applied to the two grid electrodes creating a high field region in the silicon pillars. The amplification is achieved in the pillars and charges drifting in this region induces a signal on the readout electrodes. The amplification electrodes can be biased either via bonding in a dedicated etched area, or through metallisation running along the pillar wall in one of the etched region.\\
Deep reactive ion etching process (DRIE) followed by successive deposition and patterning of dielectric and metal electrode is a possible approach to the making of this device. Constraints associated to this fabrication technique are considered in the following, in particular the interplay between the depth of the etch, the thickness of the dielectric layers that will be deposited on the wall and the distance needed between the pillar wall and the metal electrodes to ensure no connection is left between the top and bottom of the pillars. Alternative fabrication approaches may lift such limitations. Single buried electrode structures may also be envisaged 
and will be discussed in section~\ref{sec:discussion}
\subsection{Geometry description}
\label{sec:description:geometry}

\begin{figure}[!h]
    \centering
    \includegraphics[width=0.99\textwidth]{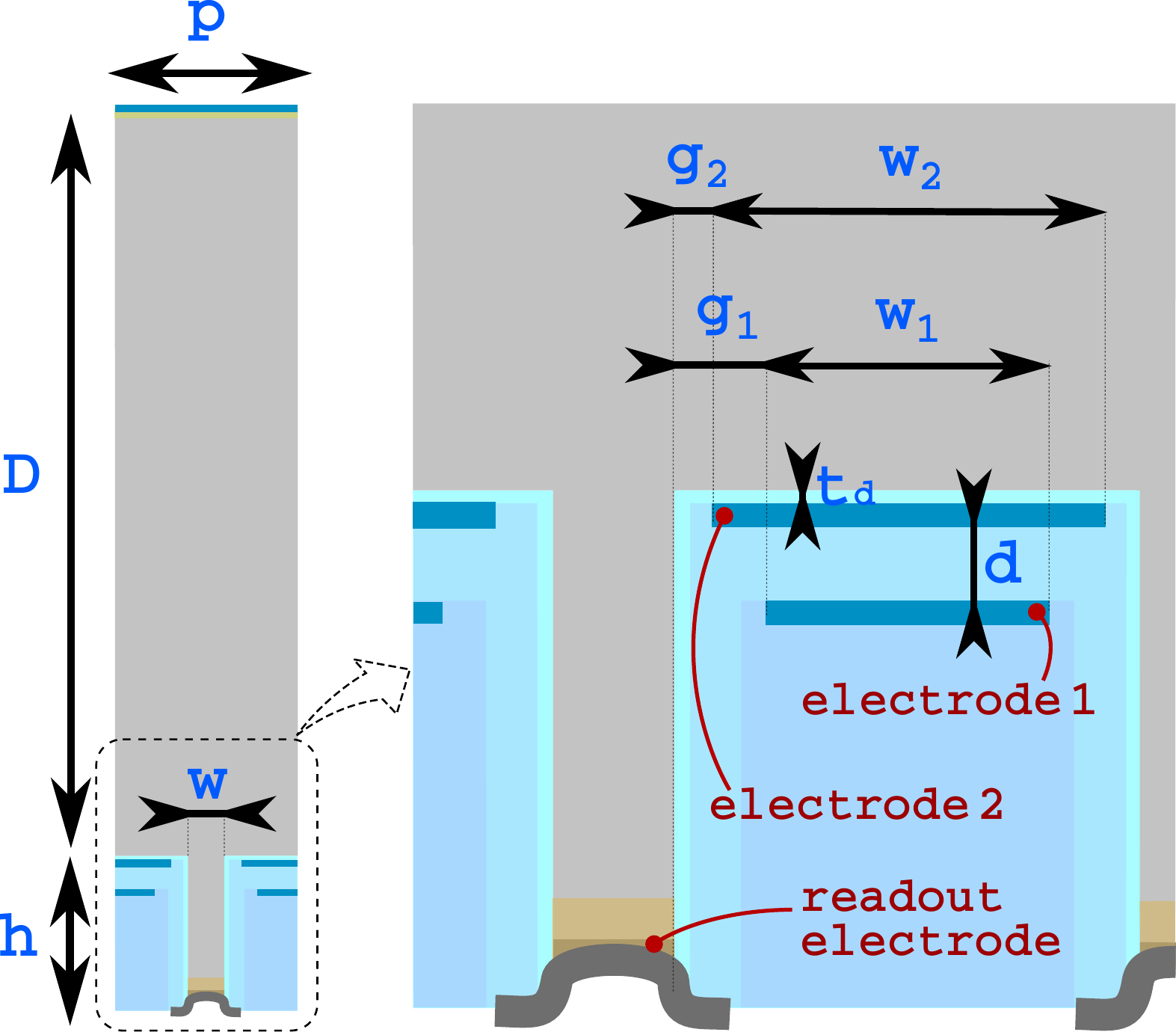}
    \caption{Definition of the key geometry parameters for the etching based fabrication process. The left schematic represents one "pillar slice".}
    \label{fig:param}
\end{figure}

The text in this section will refer to Figure~\ref{fig:param}, shows a cross-section of the SiEM design, identifying the most important dimensions.
The thickness of the drift region (D) and the pitch (p) between pillars can be optimised depending on the application. Large thickness associated with small pitch readout could be used to reconstruct the direction of the impinging particle while a thin drift region is preferred for fast timing application, in order to minimise the time distribution centroid of the signal and thus improving on time resolution~\cite{Riegler_2017}.

In the double buried electrodes geometry the electrode closest to the readout electrode is numbered \textbf{1} while the one closest to the drift region is numbered \textbf{2}.
The pillar width (w), the pillar height (h), and the pillar pitch (p), as well as the electrodes geometry, their widths (w$_1$, w$_2$) and the inter-electrode distance (d) are the principal drivers which define the electric field shape and its intensity. A few other geometrical parameters are shown in Figure~\ref{fig:param} and are related to the expected process limitation. The distances g$_1$ and g$_2$ are the guard distances between the electrodes and the pillar wall while t$_d$ designate the thickness of the dielectric between the bottom of the bulk and the \electrodetwo which is the first electrode to be deposited in the considered fabrication process.\\
In the following sections the behaviour of the device and the impact of these dimensions on the behaviour of the device is investigated using simulations.

\section{Sensor behaviour}
\label{sec:simulation}
\subsection{Setup}
\label{sec:simulation:setup}

The TCAD device simulator Synopsis Sentaurus~\cite{tcad,sentaurus} version S-2021.06 has been used to perform the simulation of the proposed structure. The 2D-mesh has been made by the Sentaurus Structure Editor with mesh cell sizes ranging from 200 to \SI{10}{\nano \meter}. The PARDISO solver~\cite{pardiso-7.2d,pardiso-7.2e} along with the Canali mobility model~\cite{canali}, Slotboom band gap model~\cite{SLOTBOOM_1, SLOTBOOM_2, SLOTBOOM_3}, and the vanOverstreaten - de Man impact ionization model~\cite{vanoverstreaten} have been applied for quasi-static simulations. In transient simulations, the HeavyIon charge deposition is used.

Unless otherwise specified, the values for the geometrical dimensions defined in \ref{sec:description:geometry} are the following: the silicon parameters are defined by D=\SI{50}{\micro \meter}, p=\SI{10}{\micro \meter}, h=\SI{6}{\micro \meter} and w=\SI{2}{\micro \meter} while the electrode geometry is given by d=\SI{1}{\micro \meter} and t$_d$=\SI{0.5}{\micro \meter}. A slice of the sensor corresponding to one single pillar, as shown on the left of Figure~\ref{fig:param}, is referred to a "cell".

The geometry description in simulation reflects the expected constraints coming from the expected production process. In particular it is expected that when a layer of silicon oxide is deposited in the trenches, 60\% of the oxide thickness is also deposited along the wall. It is also expected that to prevent metal remaining on the pillar walls, a guard of at least \SI{0.25}{\micro \meter} is introduced between the edge of the electrodes and the pillar wall. The minimal value of g$_{2}$ depends  on t$_d$, while the minimal value of g$_1$ depends both on d and t$_d$.

In the default configuration the \electrodeone is as large as possible, i.e.\ w$_1$=\SI{5.7}{\micro \meter} while the \electrodetwo is retracted from its maximal size in order to have w$_2$=w$_1$ and thus provide a simple to interpret field geometry to start with.\\
The default voltage configuration used for the electrode biasing is the following: while the readout electrode is at ground level, the potential of \electrodeone, V$_1$, is set at \SI{-5}{\volt} and V$_2$, the potential of \electrodetwo can vary to produce the electric field needed for amplification with V$_2$=V$_1$-\dv. The potential of the backside electrode V$_\text{bias}$ is constantly \SI{-30}{V} below V$_2$ to ensure the depletion of the structure. 

\subsection{Electric field and leakage current}
\label{sec:simulation:fieldncurrent}
\begin{figure}[!ht]
    \centering
        \includegraphics[width=\textwidth]{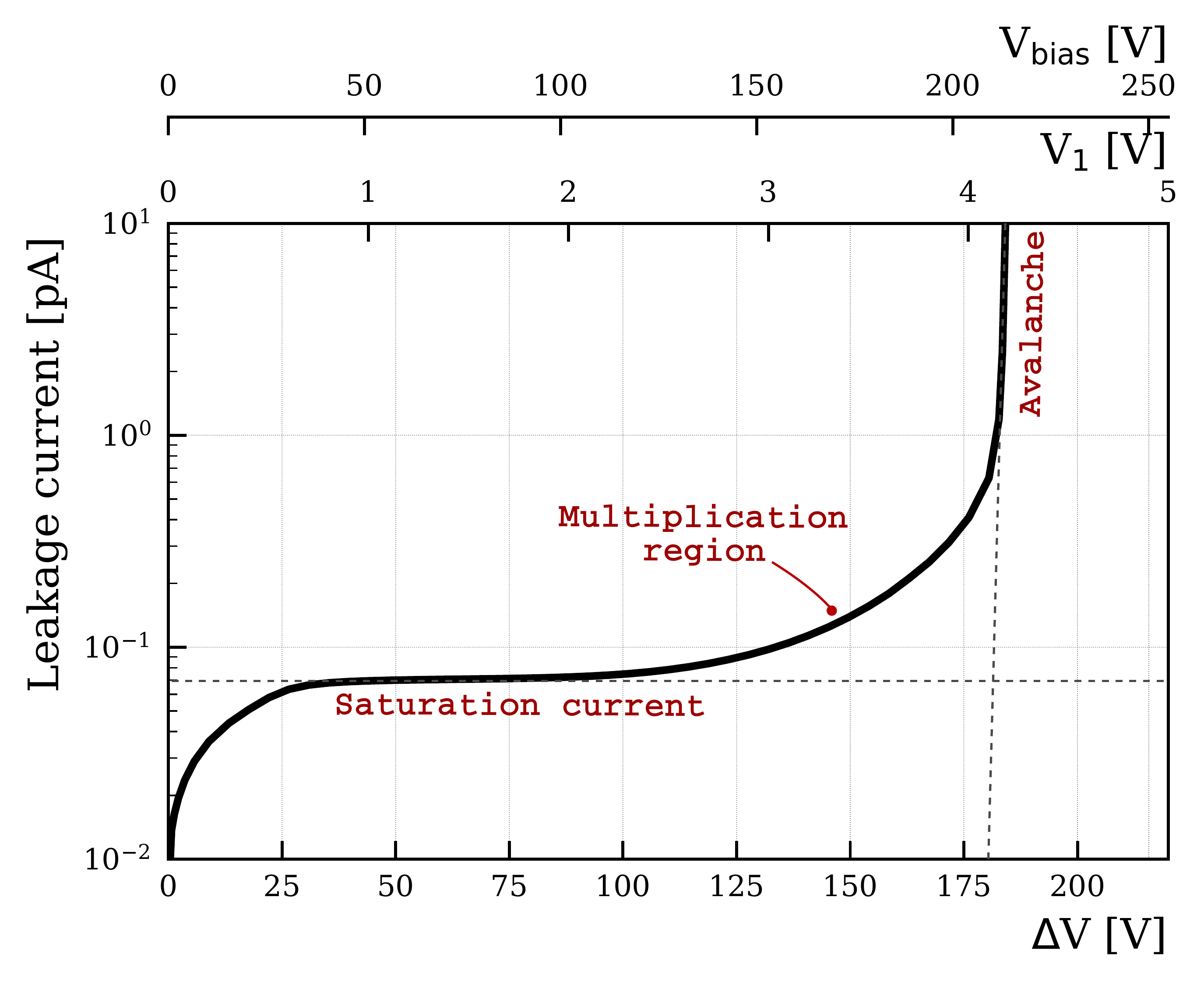}
    \caption{I-V characteristic of the device with default geometry and biasing scheme. The current is estimated for a cell which correspond to a volume of \SI{512}{\micro\meter^{3}}}
    \label{fig:ivcurve}
\end{figure}

\begin{figure}[!ht]
    \centering
    \subfloat[]{\includegraphics[width=0.47\textwidth]{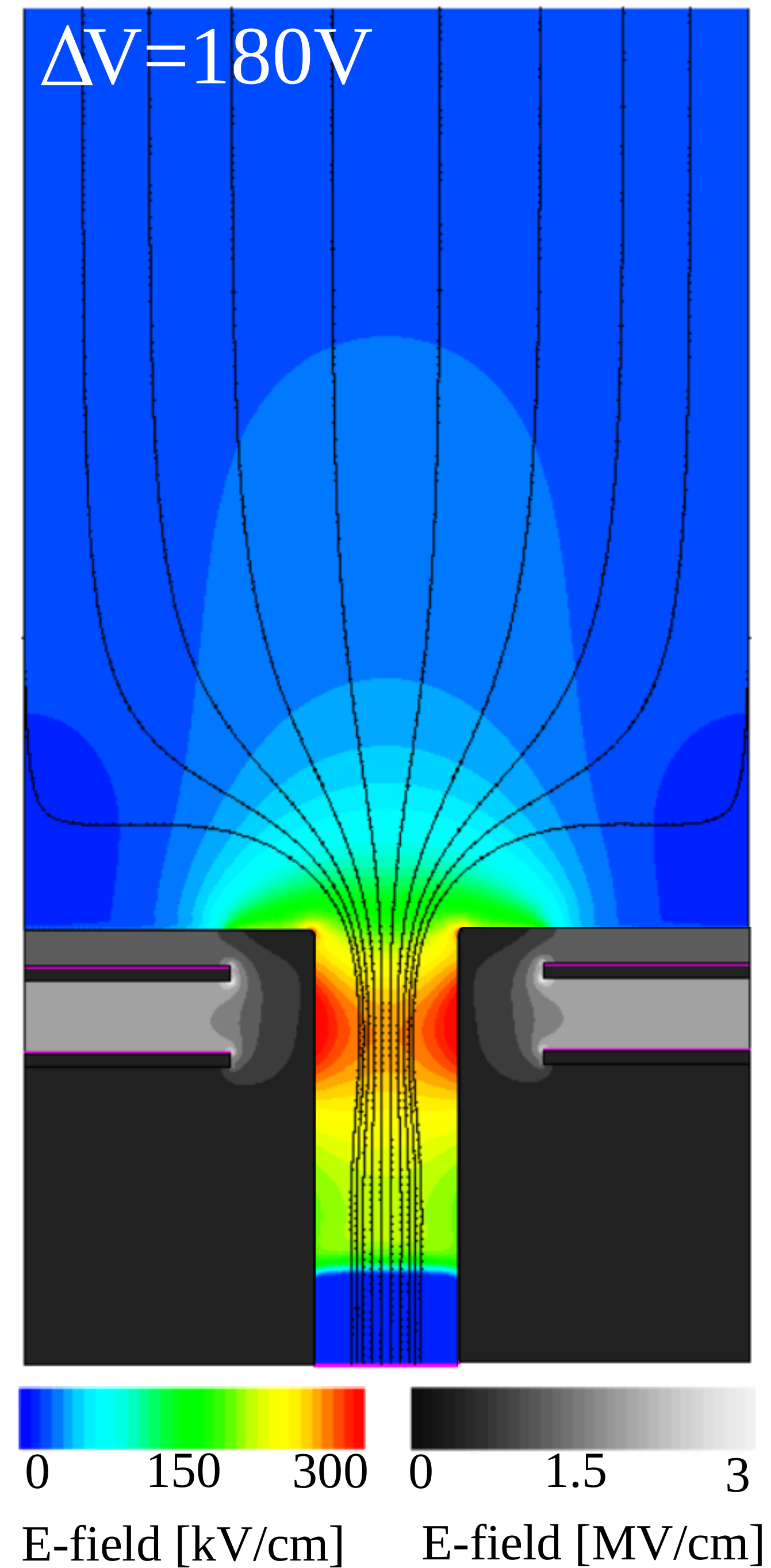}\label{fig:fieldlines:efield}}
    \subfloat[]{\includegraphics[width=0.47\textwidth]{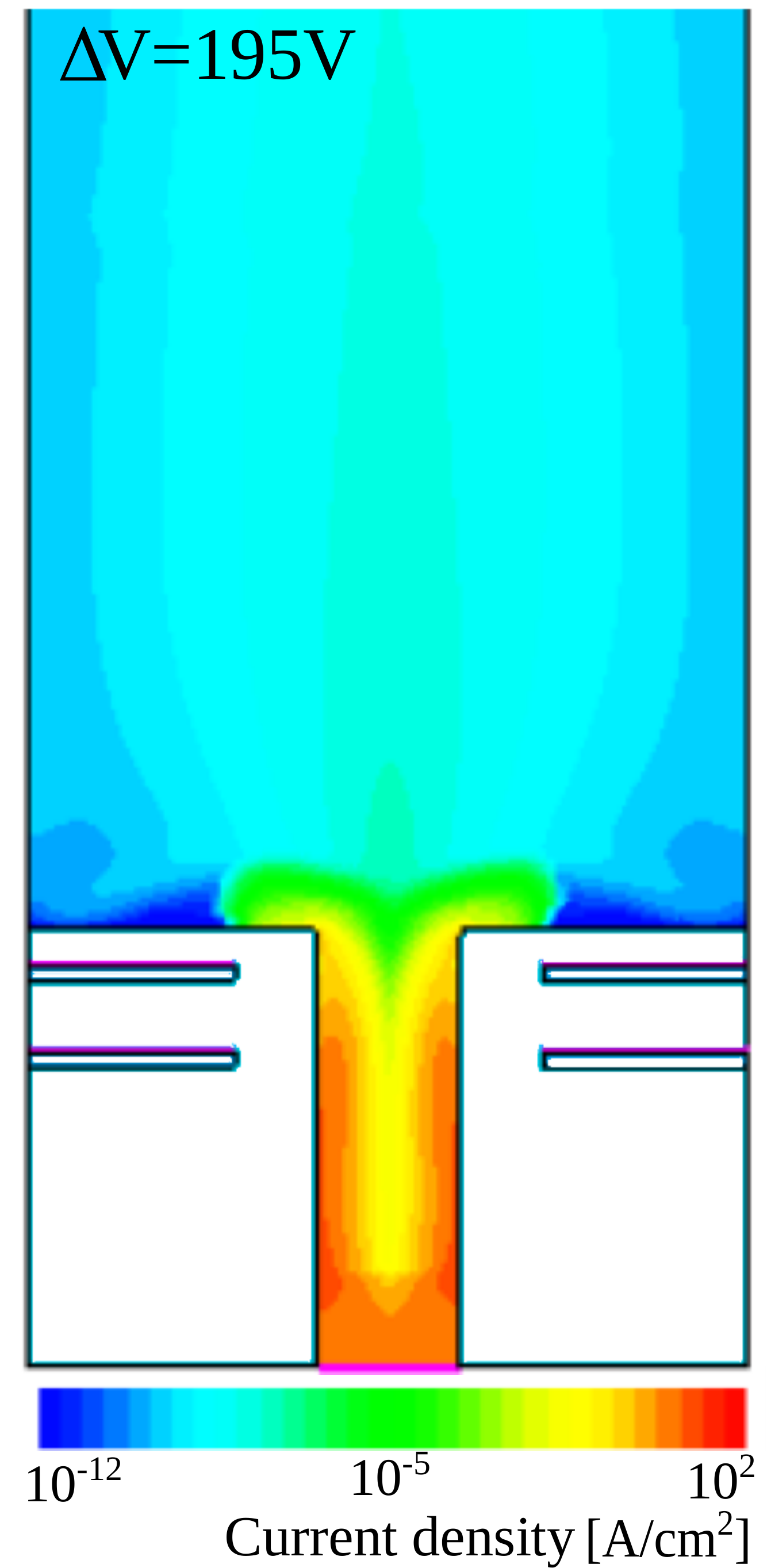}\label{fig:fieldlines:impactionisation}}\quad%
    \caption{ Maximum electric field amplitude reached just before the sensor enters breakdown (\dv=\SI{180}{\volt}) (a) and the current density just after breakdown (\dv=\SI{195}{\volt}) (b). On figure (a) the black lines represent the electron velocity streamlines.}
    \label{fig:fieldlines}
\end{figure}

In Sentaurus TCAD quasi-stationary simulations, the voltages V$_1$, V$_2$ and \vbias are ramp up simultaneously and linearly. During the voltage sweep, the leakage current at each step is evaluated, leading to a characteristic curve similar to the I-V of a planar sensor, even though the device is not bipolar. 
When the target voltages are reached, the electric field and other figures of merit discussed in \ref{sec:simulation:optimisation} are evaluated.
Figure~\ref{fig:ivcurve} shows the I-V characteristic of a cell for a \dv ramp up to a potential of \SI{235}{\volt}.
At low voltages, the I-V curve follows a reversed biased diode-like shape, until it reaches the saturation leakage current. At this point both the pillar and the drift region are depleted and the leakage current is low enough to allow primary charge particle detection ($\sim0.1$pA per cell). As \dv increases, the field eventually gets large enough for the creation of secondary charges through impact ionisation. The leakage current increases as the thermal electrons get multiplied in the pillar. In this regime high field is concentrated at the top of the pillar as illustrated in Figure \ref{fig:fieldlines}. Values of the electric field between 200 and \SI{300}{\kilo \volt/\cm} can be achieved in the silicon while keeping the field in the silicon oxide below \SI{3}{M\volt/\cm}. 

For the default pillar width of \SI{2}{\micro \meter}, above \dv of \SI{200}{\volt} a sustained avalanche develops along the pillar wall, where the field is the highest. This avalanche, which leads to the breakdown of the device, is illustrated in Figure~\ref{fig:fieldlines:impactionisation}.

\begin{figure}[!ht]
    \centering
        \includegraphics[width=0.99\textwidth]{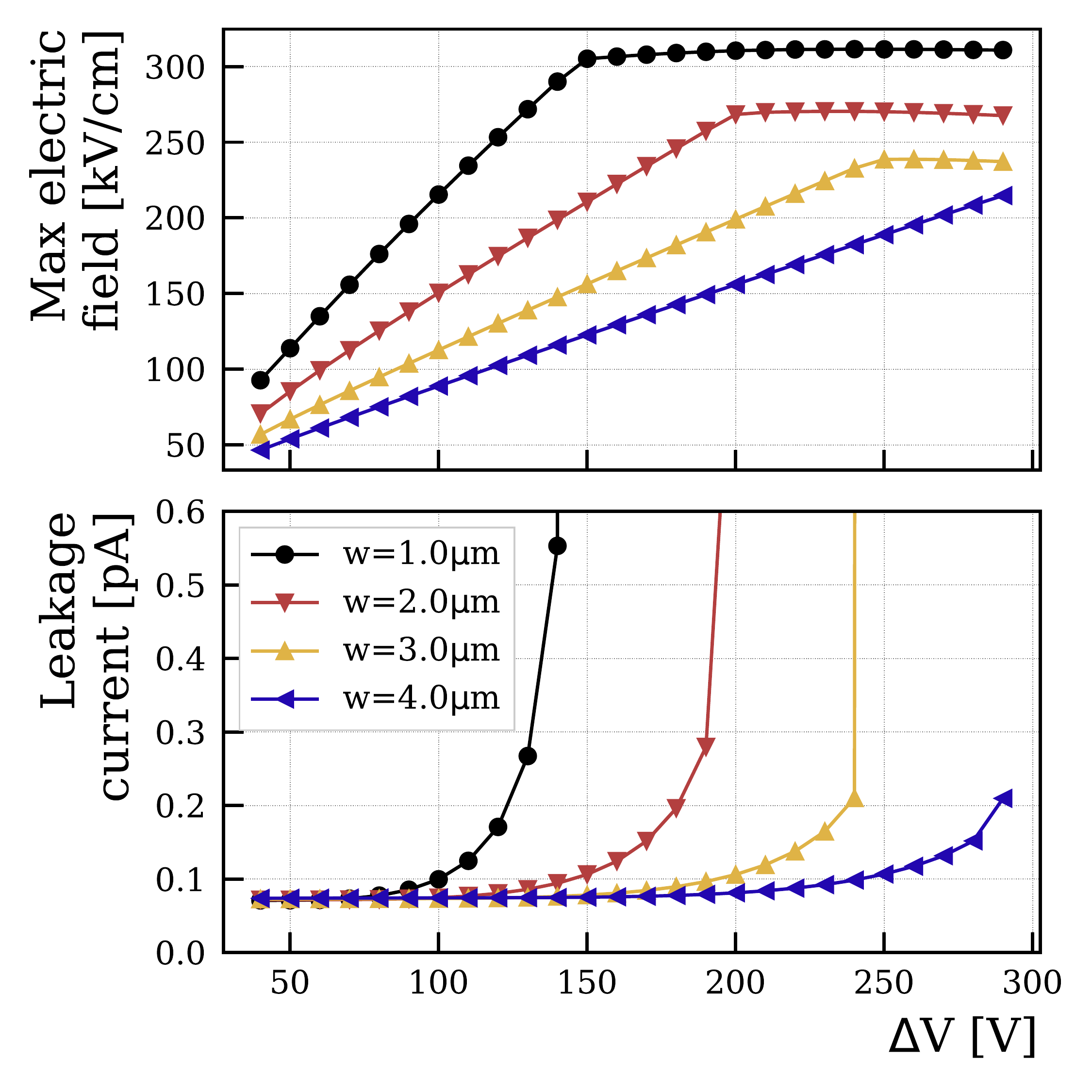}
    \caption{Leakage current and maximal electric field at the center of the pillar as function of \dv for different pillar width. Saturation of the electric field amplitude is due to polarisation effects that occurs when current flows in breakdown conditions.}
    \label{fig:field_center}
\end{figure}

The role played by w, the pillar width, was investigated and is illustrated  in Figure~\ref{fig:field_center}.  In these simulations V$_1$ is kept constant, while \dv is varied for different pillar widths. The I-V curves display similar behaviour to the one of Figure~\ref{fig:ivcurve}, but a higher \dv is required to reach charge multiplication for thicker pillars.  This is due to the fact that for the same potential difference applied to the amplification electrodes, a lower electric field amplitude is achieved in the center of the pillar. Furthermore, the electric field just before breakdown voltage is reached is lower for thicker pillars since the breakdown is initiated at the pillar wall where the field is the highest. This reduces the expected maximum amplification capacity of thicker pillars. With the DRIE based process, widths of down to w$=$\SI{0.6}{\micro \meter} can most likely be achieved. A pillar size of \SI{2}{\micro \meter} is used in the rest of the studies reported here. Note that, larger pillars are also of interest since, as will be discussed in \ref{sec:simulation:optimisation}, other parameters than the maximum electric field in the center of the pillar are relevant to the amplification mechanism.
\subsection{Signal amplification}
\label{sec:simulation:amplification}
\begin{figure}[h!]
    \centering
    \subfloat[]{\includegraphics[width=0.95\textwidth]{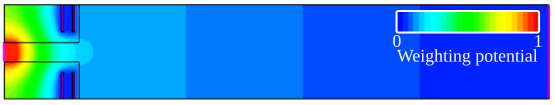}\label{fig:w_pot:top}}\quad
    \subfloat[]{\includegraphics[width=0.95\textwidth]{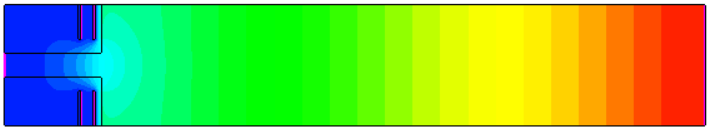}\label{fig:w_pot:bottom}}\quad
    \caption{Weighting potentials of the readout electrode (a) and backside electrode (b). The representation is rotated by 90$^{\circ}$ with respect to the other figures.}
    \label{fig:w_pot}
\end{figure}

In the previous paragraph it was shown that fields above \SI{200}{\kilo \volt/cm} can be created in the pillar without breakdown of the device. Primary charge should thus be multiplied by impact ionisation in the pillar, providing internal gain to the sensor. Transient simulations are used to probe the multiplication mechanism. A charge cloud of 80 electron--hole pairs is deposited in the drift region center, and propagated in the device until all charges are collected at the electrodes.
During the charge transport, the induced currents at the electrodes are recorded and further processed in order to extract the gain and the signal shape. The gain is defined as the ratio of integrated current on the readout electrode by the input charge. The integrated charge at the readout electrode is leakage current subtracted.
 \begin{figure}[t]
    \centering
    \subfloat[]{\includegraphics[width=0.95\textwidth]{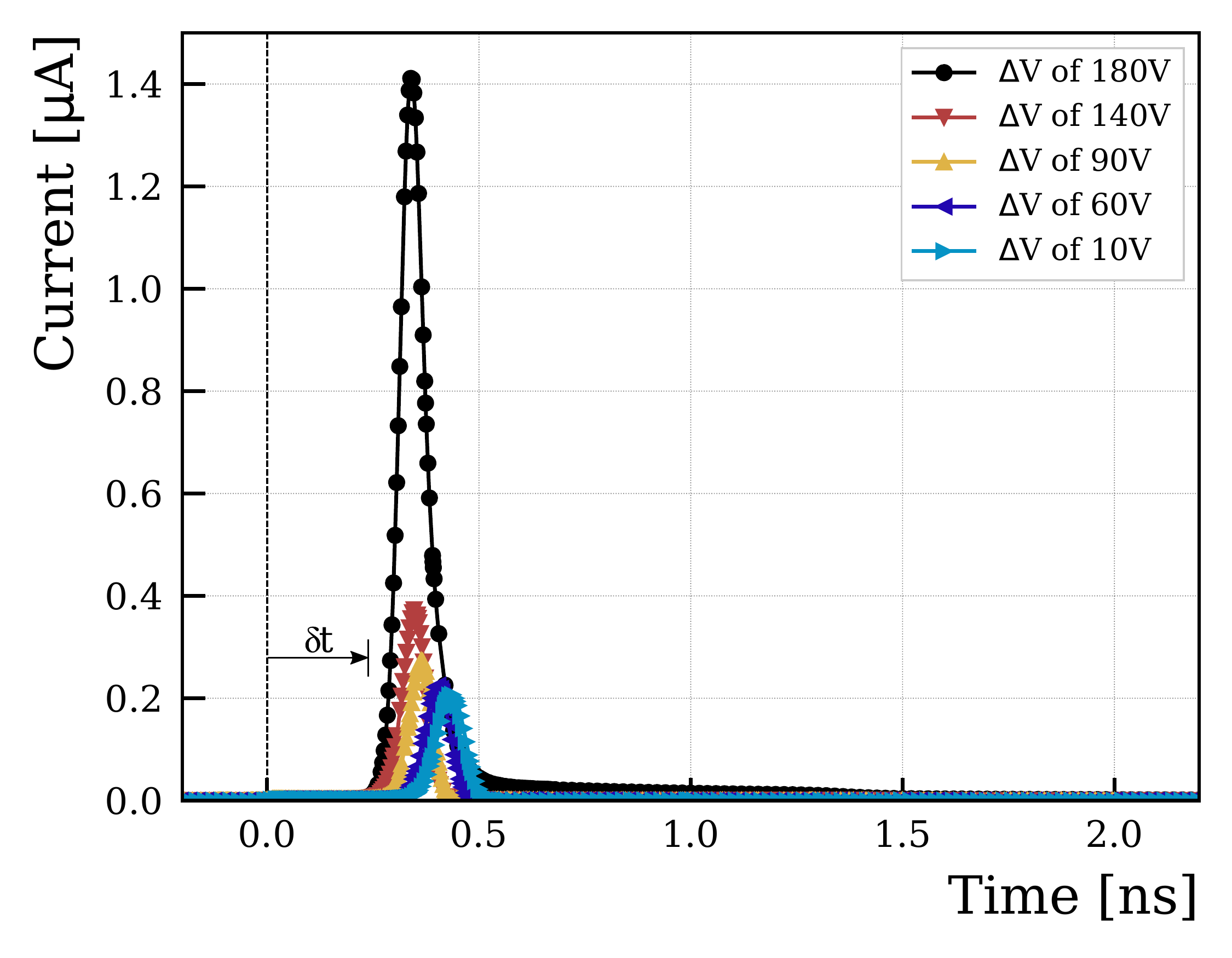}
    \label{fig:trans_currents:readout}}
    \quad
    \subfloat[]{\includegraphics[width=0.95\textwidth]{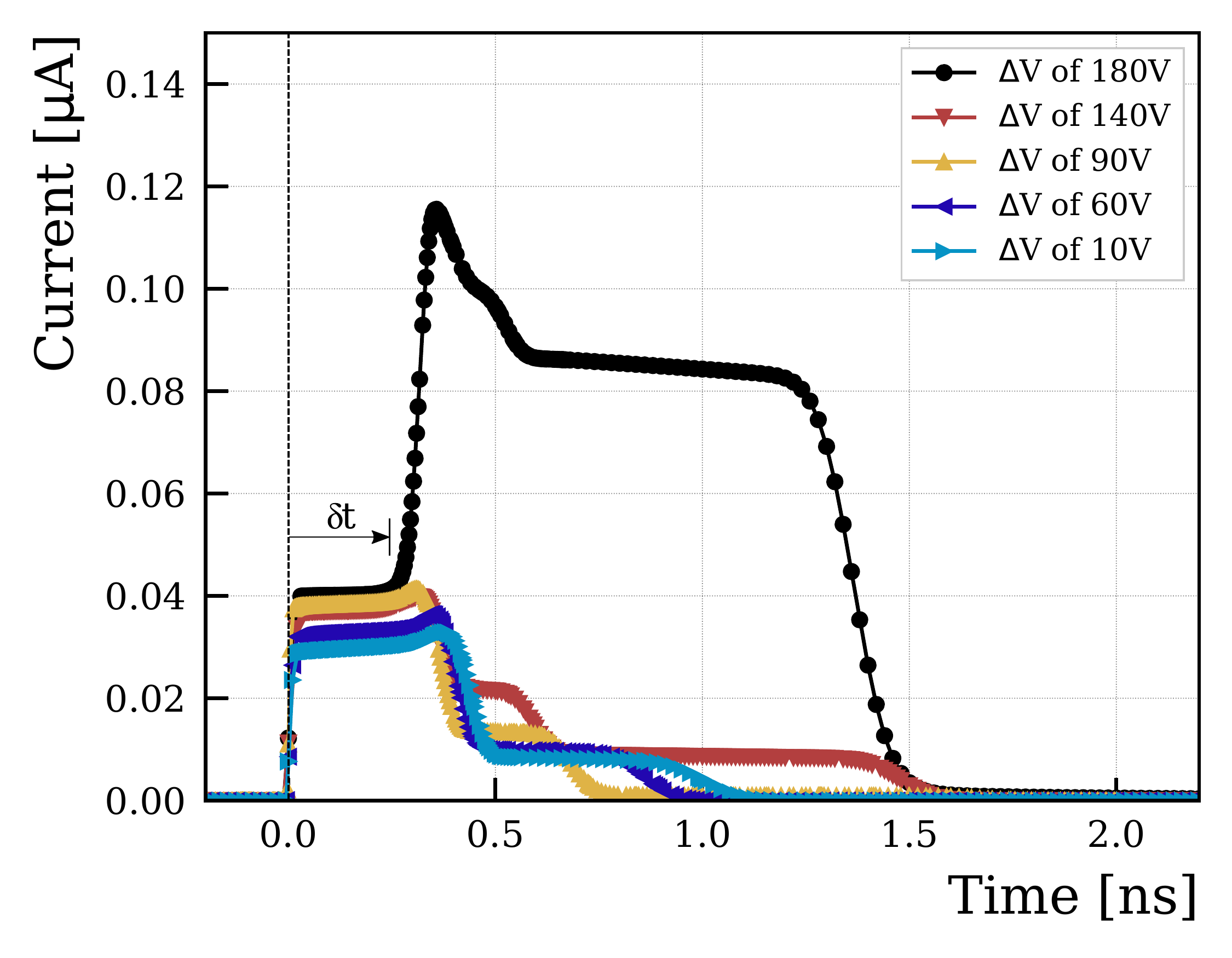}
    \label{fig:trans_currents:backside}}
    \quad
    \caption{Current induced on the readout electrode (a) and the backside electrode (b) for \dv between \SI{10}{\volt} and \SI{160}{\volt}. $\delta$t indicates the time needed for the electrons to reach region \induction for a \dv of \SI{160}{\volt}.}
    \label{fig:trans_currents}
\end{figure}
In Figure \ref{fig:trans_currents:readout}, the currents from transient simulations using different \dv are displayed. Although the signal induced at the readout electrode in Figure~\ref{fig:trans_currents:readout} is the one of interest for MIP detection, the signal induced on backside electrode is interesting to understand the signal formation and is show in Figure~\ref{fig:trans_currents:backside}.\\
At \dv=\SI{10}{\volt} the device exhibits no gain. On the backside electrode the primary holes induce a signal from t=0~ns until all holes have reached the electrode. 
This signal mainly induced by the holes drifting towards the backside electrode looks very much like what would be observed on a pad sensor, as suggested by the distribution of the weighting potential of the backside electrode shown on Figure~\ref{fig:w_pot:bottom}. 
On the read-out electrode however, only the charge moving in the pillar induces signal. The signal only start when the electrons reach the buried electrodes. Charges moving in the drift region are shadowed by the amplification electrode as expected from the low weighting potential variation seen in Figure~\ref{fig:w_pot:top}.\\
At \dv$>$\SI{100}{\volt}, when the electrons reach the top of the pillar, where the electric field is the highest, they generate secondary charges. Secondary holes drift back to the readout electrode inducing the second wave of signal starting at $\delta$t. The amplitude of this second wave depends on the amount of charge multiplication.
The gain can be clearly seen from looking at the amplitude of the signal induced on the readout electrode, where the total secondary electron signal is added to the one from the primary electrons.
In this configuration, the gain vary from $\sim$2 for \dv=\SI{140}{\volt} to slightly above 10 for \dv=\SI{180}{\volt}. It can be visualised on the Figure~\ref{fig:trans_currents:readout} by comparing the area under the curves at \dv=\SI{140}{\volt} and \dv=\SI{180}{\volt} with the area under the \dv=\SI{10}{\volt} in which the field is too low for multiplication to occur.

 \begin{figure}[!ht]
\centering
  \subfloat[]{\includegraphics[width=0.48\textwidth]{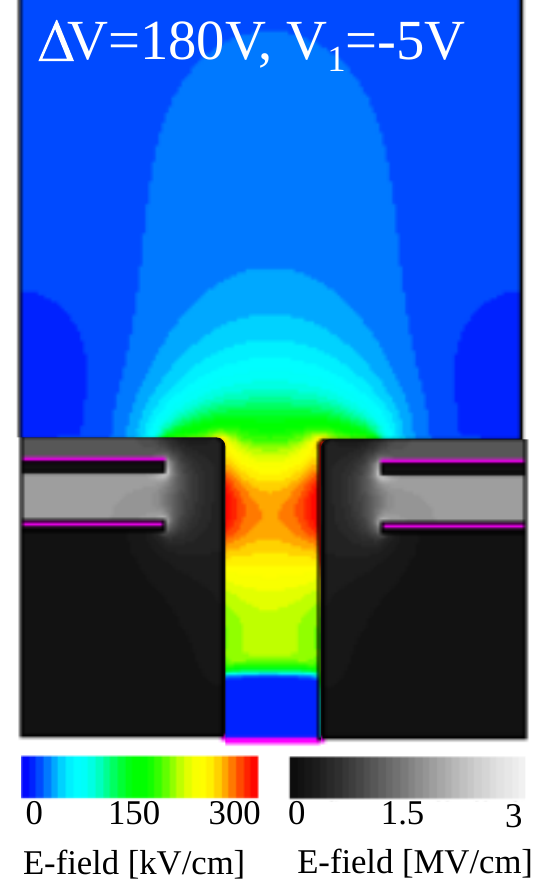}\label{fig:fieldmap:gem}}
  \subfloat[]{\includegraphics[width=0.48\textwidth]{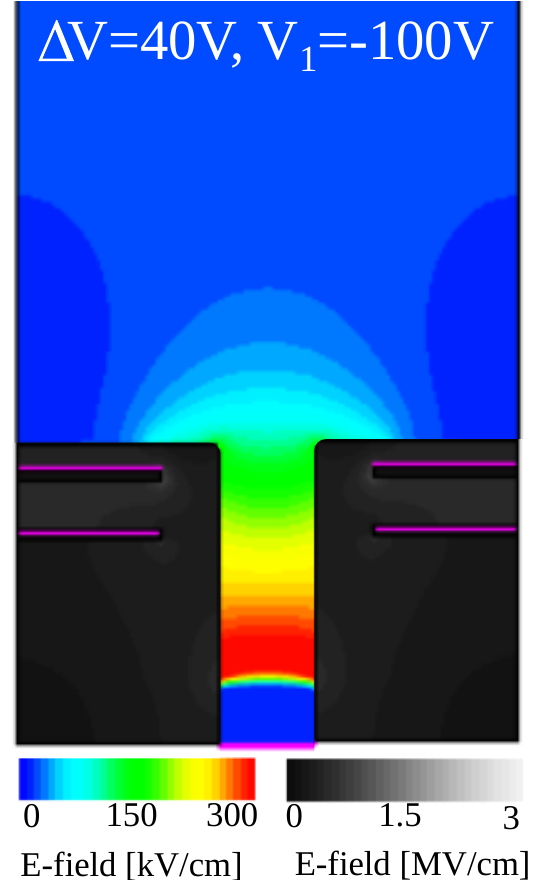}\label{fig:fieldmap:single}}\quad
\caption{Electric field in pillar for different biasing configurations}
\label{fig:bias_config}
\end{figure}

\subsection{Optimisation of the device}
\label{sec:simulation:optimisation}
\begin{figure*}[!ht]
\centering
  \subfloat[]{\includegraphics[width=0.49\textwidth]{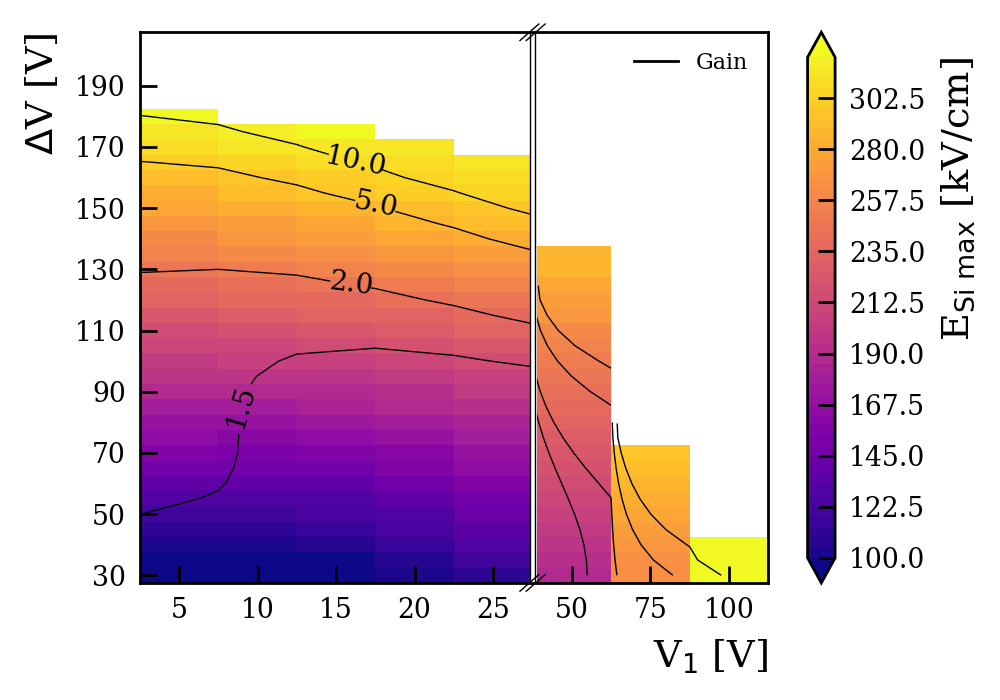}\label{fig:vdvscan:simax}}
  \subfloat[]{\includegraphics[width=0.49\textwidth]{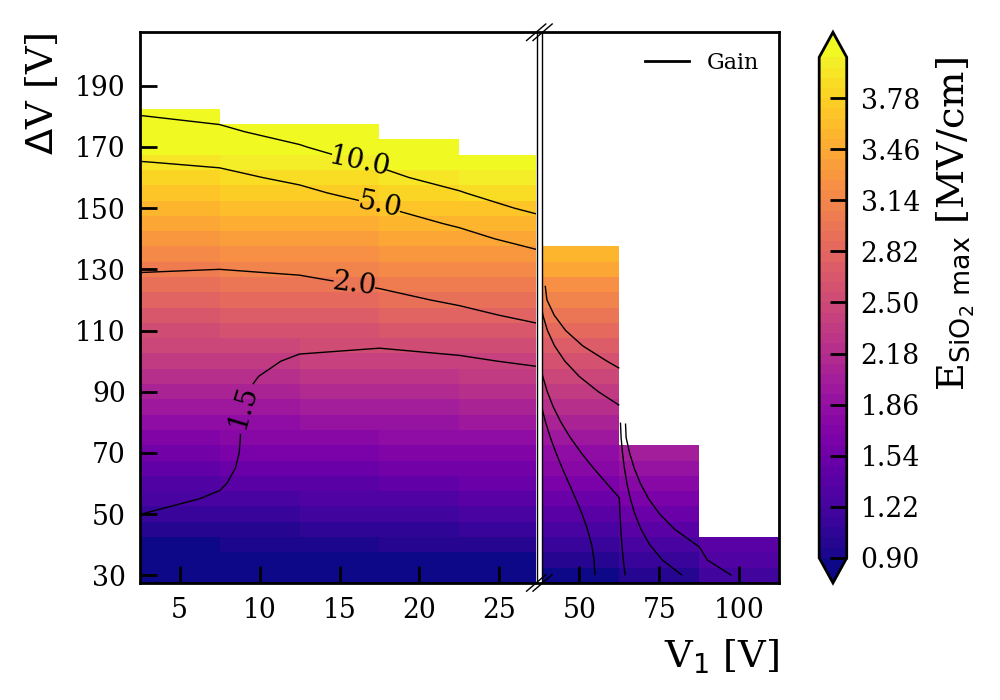}\label{fig:vdvscan:sio2max}}\quad
  \subfloat[]{\includegraphics[width=0.49\textwidth]{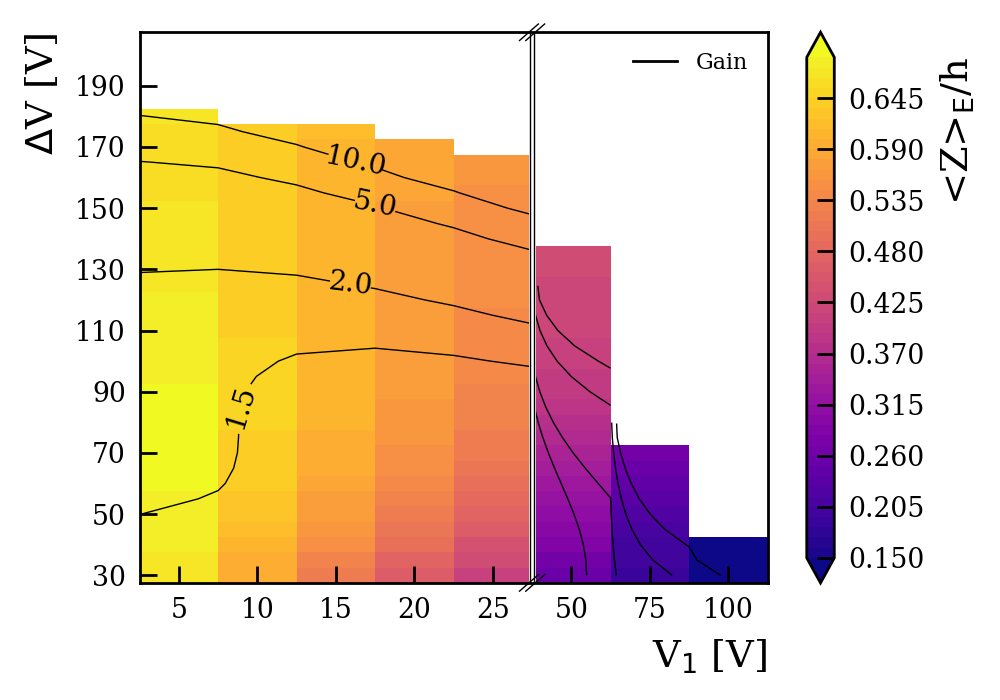}\label{fig:vdvscan:bary}}
  \subfloat[]{\includegraphics[width=0.49\textwidth]{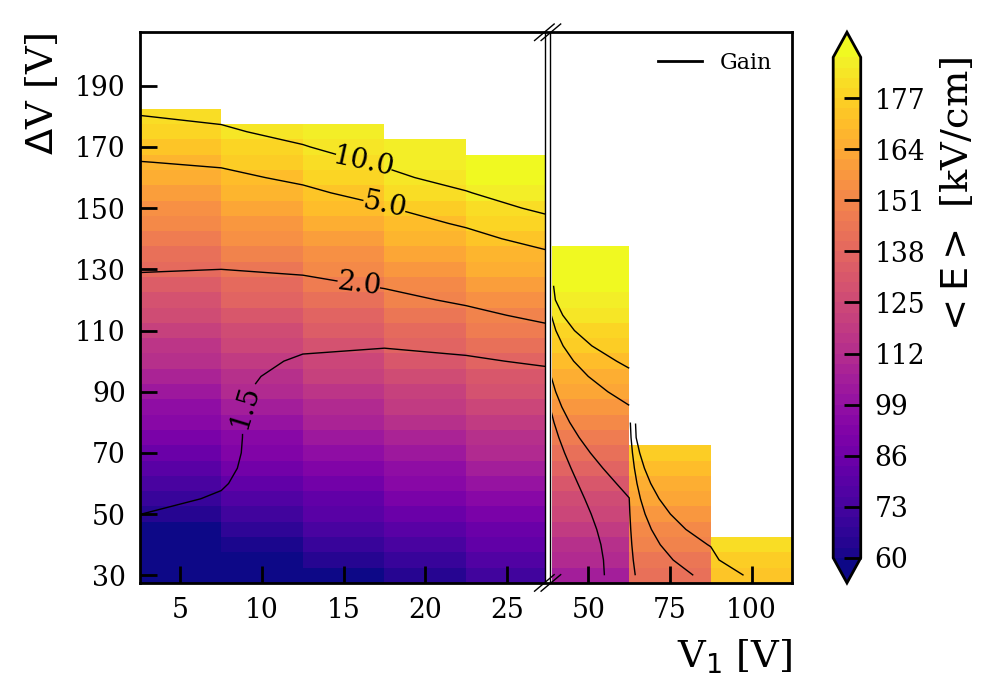}\label{fig:vdvscan:avergae}}%
  \caption{The maximum field in silicon and silicon oxide,  E$_{\mathrm{Si~max}}$ (a), and E$_{\mathrm{SiO_2~max}}$ (b) evolution in the (\vone,\dv) phase-space is presented, together with the evolution of the field barycenter, $<\mathrm{Z}>_{\mathrm{E}}$/h (c) and the average value of the field in the pillar, $<\mathrm{E}>$ (d). The gain is overlaid in each figure. Simulation were performed on about 60 points of the (\vone,\dv) phase space and cubic 2D-interpolation is used to obtained a continuous representation.}\label{fig:vdvscan}
\end{figure*}

In the previous paragraphs, the configuration, both in terms of electrode geometry and in terms of biasing scheme is close to that of a typical GEM detector~\cite{gem}, with the high field region located in the interstice of the multiplication electrodes structure as illustrated in Figure~\ref{fig:fieldmap:gem}. In the following, configurations where the high field region goes further into the pillar are investigated. In particular, high field can be generated by a difference of potential between \electrodeone and the readout electrode. This configuration could be achieved with a single electrode and, as illustrated in Figure~\ref{fig:fieldmap:single}, the highest field is shifted towards the readout electrode. A well chosen balance between those two configurations would, on one hand, prevent generation of a localised high field which could initiate breakdown either in the silicon or in the silicon oxide and, on the other hand, increase the fraction of the pillar volume where charge multiplication can occur.
This can be achieved by changing the electrode geometry or by modifying the biasing scheme, as discussed in the following paragraphs.

\paragraph*{Figure of merit} In general, gains of above a factor 10 are targeted. It is nonetheless useful to define additional variables in order to understand the advantages or limitations of each configuration. The maximum electric field in the silicon, E$_{\mathrm{Si~max}}$, and in the silicon oxide,  E$_{\mathrm{SiO_2~max}}$, indicate if a configuration is close to a breakdown conditions. High values of  average electric field in the pillar, $<\mathrm{E}>$ reflect higher field achieved over a larger region of the pillar, especially when E$_{\mathrm{Si~max}}$ is not large. Finally the barycenter position of the electric field along the pillar axis,\bary , give information on the position of the high field and how localised the field is. The closer it is to one electrode the more peaked it is, the closer to the pillar center, the more spread along the pillar it is. For this variable not to depend too much on the geometry, it is normalised by the height of the pillar, h. Ideal configurations have high $<\mathrm{E}>$, a medium \bary and as low E$_{\mathrm{Si~max}}$, and E$_{\mathrm{SiO_2~max}}$ as possible.

\paragraph*{Interplay between V$_1$ and \dv}

\begin{figure}[h!]
\centering
  \includegraphics[width=0.99\textwidth]{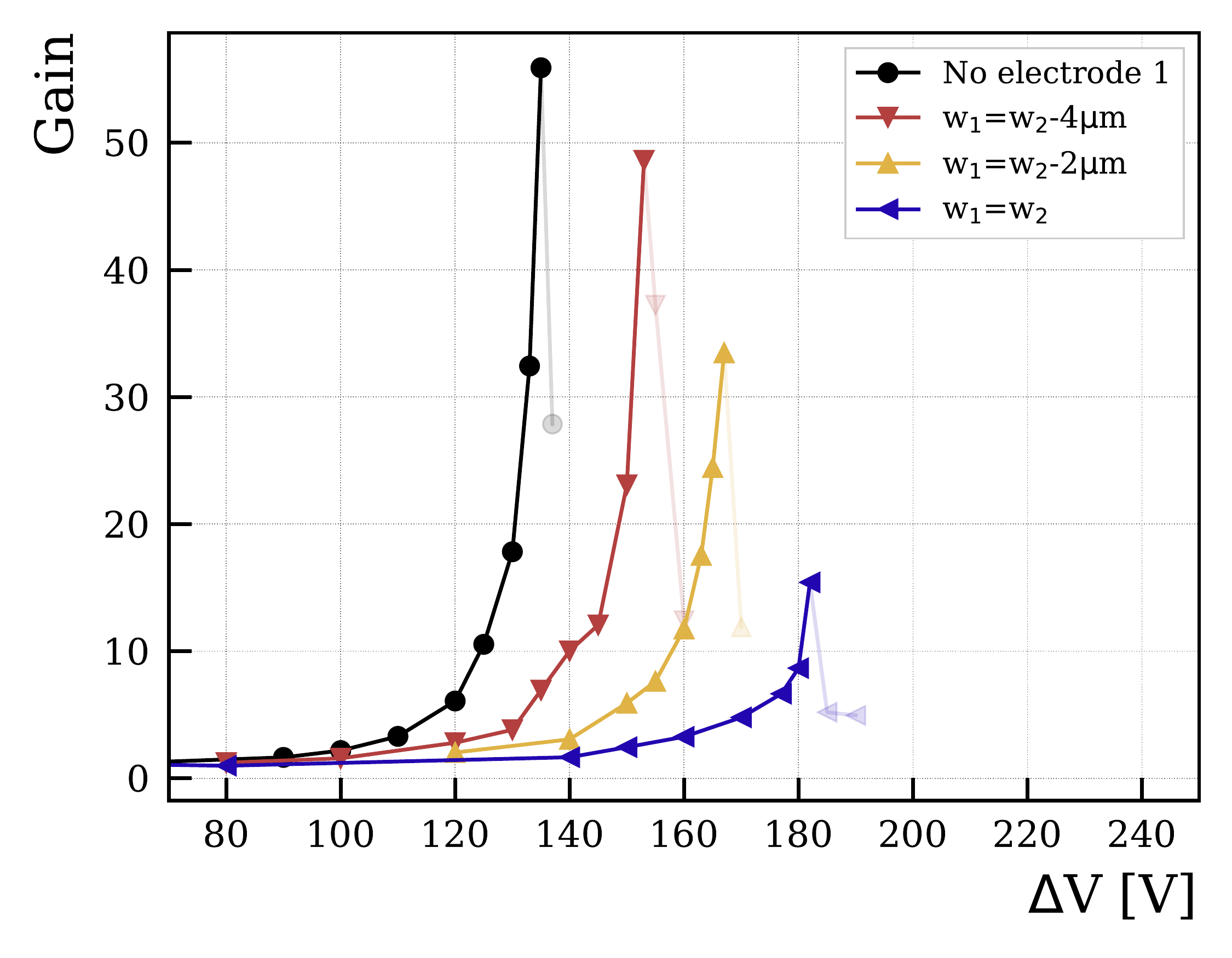}
\caption{Gain as function of \dv for different widths of \electrodeone. The \electrodeone width is downsized from the default w$_1$=w$_2$=\SI{5.7}{\micro \meter} down to its complete removal. The points in light color correspond to values of \dv where the structure is in breakdown.}
\label{fig:retraction}
\end{figure}

\begin{figure}[h!]
\centering
  \includegraphics[width=0.99\textwidth]{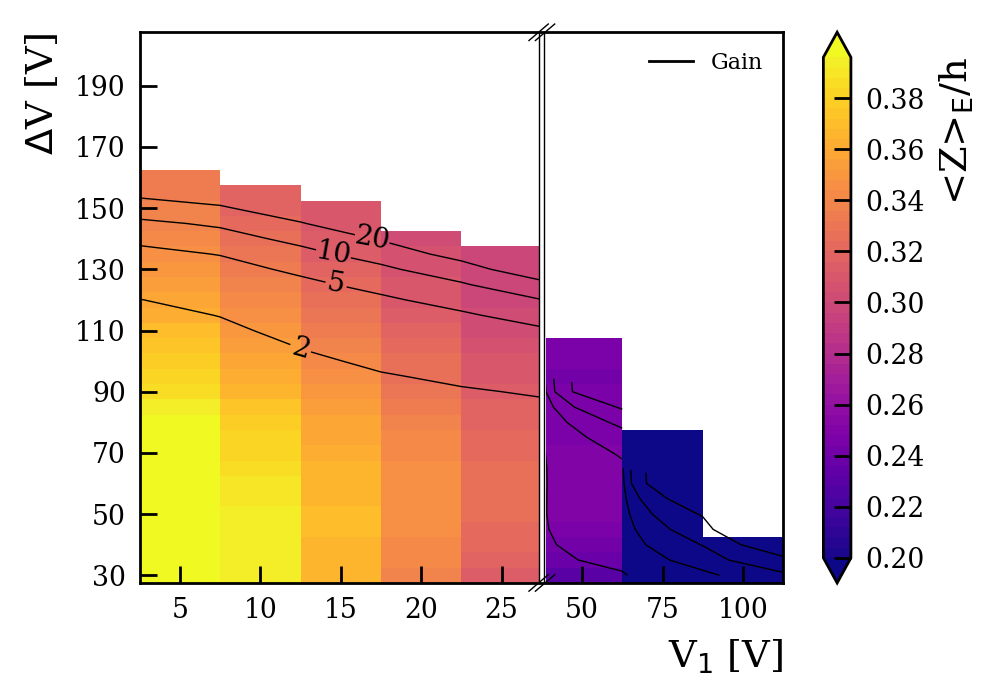}
\caption{The evolution of the field barycenter, \bary/h and the gain as function of the biasing configuration (\vone,\dv) for a geometry with w$_1$=w$_2$-\SI{3}{\micro\meter}. Simulations were performed on 60 points of the (\vone,\dv) phase space and cubic 2D-interpolation is used to obtain a continuous representation. }
\label{fig:retraction2D}
\end{figure}

The impact of the biasing scheme on the aforementioned figures of merit is studied in the following. 
\dv is scanned in steps of \SI{5}{\volt} until breakdown and for different values of V$_1$. The resulting gain and figures of merit are reported in Figure~\ref{fig:vdvscan}. As expected it is possible to achieve gain above a factor 10 with different balances of V$_1$-\dv, nonetheless the way amplification is achieved in the structure is very different. If \dv is large while V$_1$ is small, high electric field amplitudes are reached both in the silicon pillar, as shown in Figures~\ref{fig:vdvscan:simax}, and in the dielectric, as shown in Figure~\ref{fig:vdvscan:sio2max}. Depending on the quality of the oxide this could lead to breakdown in the dielectric. When the gain is obtained by high difference of potential between the readout electrode and the \electrodeone, ie. V$_1$ is large and \dv small, the field in the silicon reaches very high values close to the readout electrode and becomes the most likely cause of breakdown of the structure. In contrast, for V$_1$ values between \SI{50}{\volt} and \SI{75}{\volt}, for which gains of above a factor 10 can be achieved with settings of \dv above 110-\SI{60}{\volt}. These settings limit the field in the silicon below 250-\SI{300}{\kilo\volt/cm} and in the silicon oxide below  3-\SI{1.7}{M\volt/cm}) which would provide a more robust operation point further away from breakdown. The field barycenter position along the pillar height in Figure~\ref{fig:vdvscan:bary} is, as expected, the closest to the readout electrode for large V$_1$, close to the amplification electrodes in the default biasing scheme, but more central with the "robust operation" scheme which indicates that the pillar is better filled with electric field. This is further supported by the fact that at this point the average electric field shown in Figure~\ref{fig:vdvscan:avergae} is high.
\paragraph*{Electrode geometry}
A similar distribution of the regions of high electric field amplitude all along the pillar can be achieved by modifying the electrode geometry. When retracting the \electrodeone, ie. w$_1<$ w$_2$, the lower part of the pillar receives a contribution from the difference of potential between the \electrodeone and the readout electrode. In Figure~\ref{fig:retraction}, the gain as function of \dv is shown for various geometries of the amplification electrodes, from the default geometry where  w$_1$=w$_2$ to geometries with ever smaller dimension of w$_1$ ($<w_2$) down to the complete removal of  \electrodeone. The more retracted \electrodeone is, the lowest \dv needs to be in order to achieve the same gain. This is due to the fact that the smaller w$_1$ is the more of the pillar is filled with high electric field, even though the maximum field is lower.\\ 
This approach can be combined with modification to the (\vone,\dv) balance. The barycenter \bary and gain variation as function of the biasing configuration are shown in Figure~\ref{fig:retraction2D}. As expected lower \dv allow to reach the high gain region, and overall larger gains can be achieved.

\begin{figure}
        \centering
        \includegraphics[width=0.99\textwidth]{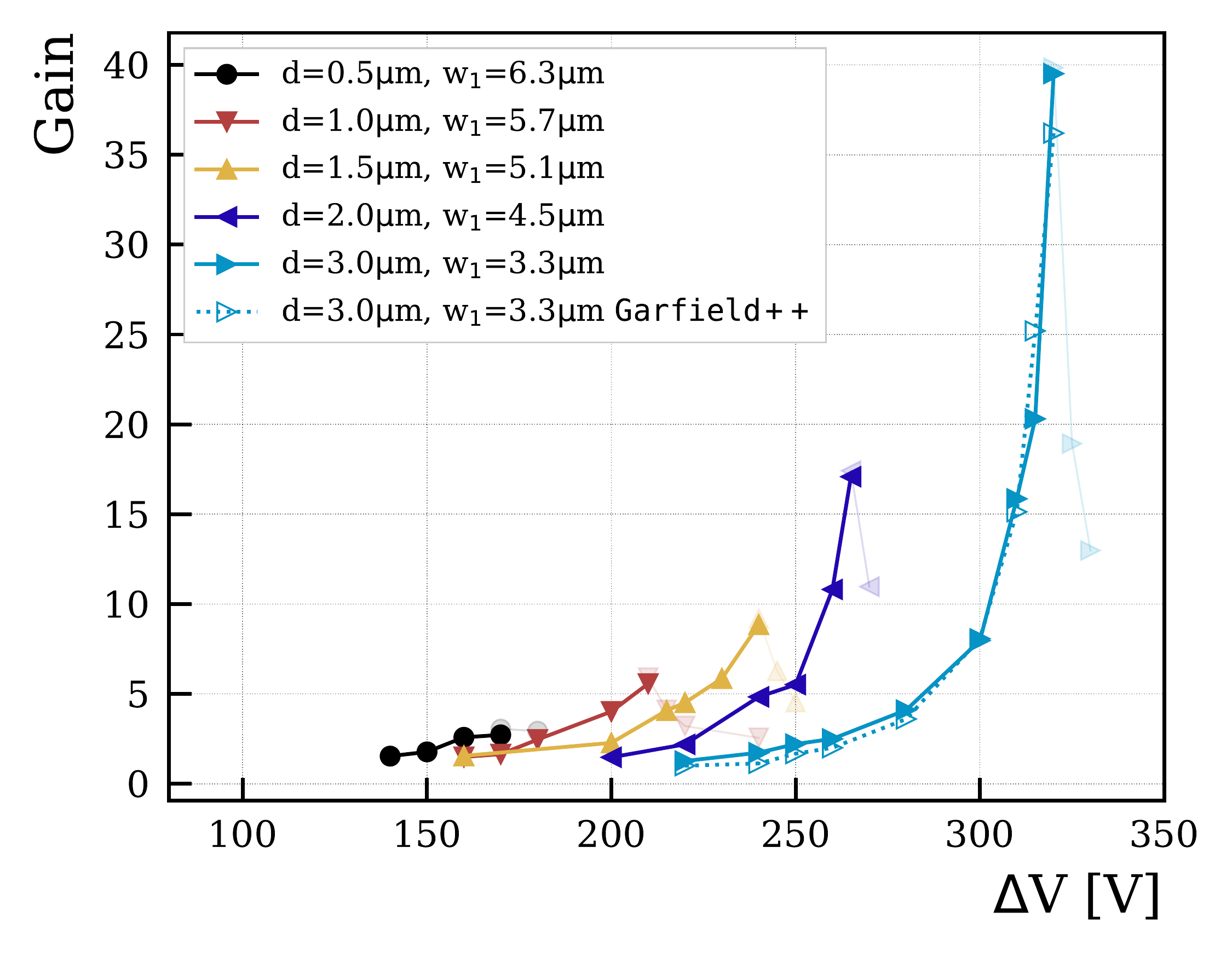}
    \caption{Gain as a function of \dv for different inter-electrode spacing d. w$_1$ varies together with d due to the expected process constraints and are reported in for completion. For the d=\SI{3}{\micro\meter} geometry, the gain obtained from alternative \texttt{Garfield++} simulation can be compared to the one obtained from Sentaurus TCAD.}
    \label{fig:gap_study}
\end{figure}

\paragraph*{Inter-electrode spacing}
In the default configuration, an inter-electrode spacing of d=\SI{1}{\micro\meter} is considered. In the DRIE based process, after the patterning of the \electrodetwo metal layer, silicon oxide is deposited before depositing \electrodeone. During this phase, about  60-70\% of the silicon oxide thickness d will also be deposited along the pillar wall. Thus, for larger values of d, g$_1$ is also larger. The variation of the gain with increasing d is studied. In order not to mix effects, and although it was shown in the previous paragraph that w$_1<$w$_2$ can be beneficial, we keep w$_1$=w$_2$ while varying d, i.e.\ w$_1$ is determined by g$_{1}$ and the \electrodetwo is shortened to have the same length than \electrodeone. The height of the pillar has been modified for each inter-electrode spacing in order to keep a constant distance, of \SI{7}{\micro\meter}, between the readout electrode and \electrodeone. This was done to limit the dependency of the study on factors other than the spacing of the amplification electrodes. The gain evolution with d is shown as function of \dv in Figure~\ref{fig:gap_study}. Since the inter-electrode distance is larger, a larger \dv is need to achieve the same gain. On the other hand the breakdown voltage is higher and the highest gain that can be achieved is larger with larger d as a bigger volume of the pillar is filled with high electric field amplitude. Gains of up to a factor 40 are observed for d=\SI{3}{\micro \meter} at \dv=\SI{320}{\volt}. Together with optimised retraction of the electrode and/or different \vone/\dv balance, higher gains are likely to be achievable.  

\subsection{Optimisation for time resolution}
\begin{figure}
        \centering
        \includegraphics[width=\textwidth]{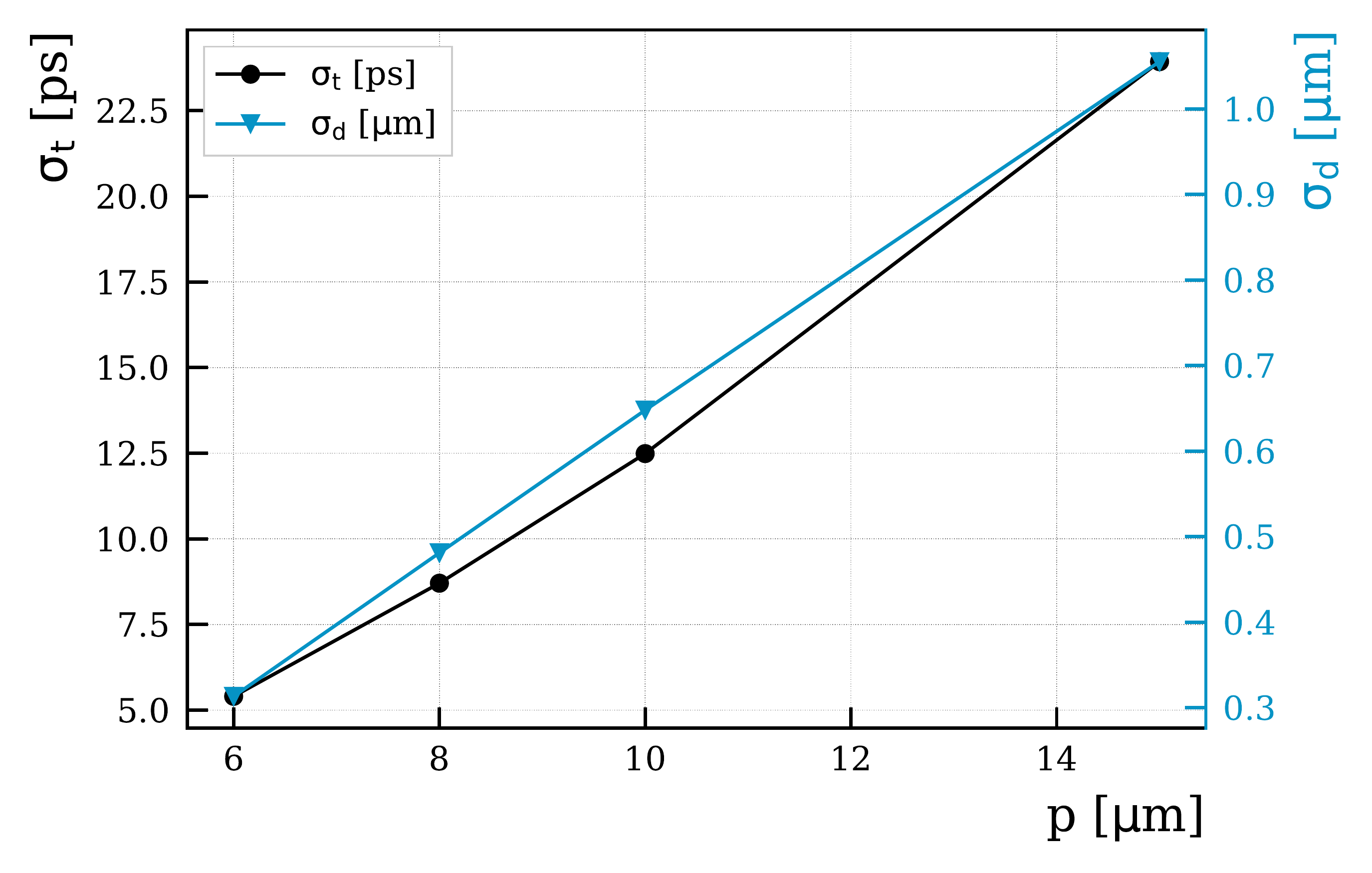}
        \caption{The standard deviation of the drift distance, $\sigma_{\mathrm{d}}$, (in blue) and of the drift time, $\sigma_{\mathrm{t}}$, (in black) between charges spread uniformly in the transverse direction in \drift and left to drift until \induction is shown as function of the pitch p.}
    \label{fig:pitch_scan}
\end{figure}

While full simulations of the deposition, drift and amplifications of primary charge carrier from MIPs, are needed to estimate the true time resolution of the sensor, the impact of some of the sensor parameters on the time structure of the signal can be assessed already. The time distribution of the signal induced on the readout electrode by the secondary electrons is driven by the time of arrival of the primary electrons at the amplification region. Following the arguments for sensor with internal gain in~\cite{Riegler_2017} and assuming the time for the secondary electrons to drift through the pillar is small\footnote{the thickness of region \induction is small with respect to the thickness of region \drift and the electron velocity in \induction is expected to be close to saturation.
}, the time resolution of SiEM will be dominated by the time it takes for electrons to drift through region \drift. Following~\cite{Riegler_2017}, the centroid time standard deviation will thus be given by $\Delta_{t}= \omega\frac{\mathrm{T_{e}}}{\sqrt{12}}$ where $\omega$ is a factor that depends on the charge deposition fluctuation and the sensor thickness ($\omega\sim$0.3 in this configuration) and $\mathrm{T_{e}}$ is the time it takes electrons to drift through \drift. For \vbias =\vtwo -\SI{30}{V}, TCAD simulation gives $\mathrm{T_{e}}\sim$ \SI{500}{\pico\second} thus $\Delta_{t}\sim$ \SI{40}{\pico\second}. But $\mathrm{T_{e}}$ can be further reduced by reducing region \drift thickness or by operating at larger electric field in \drift to increase the electron velocity. With gains of a factor 10 and above, a sensor thickness of as low as \SI{25}{\micro\meter} should provide sufficient signal for detection. For \SI{25}{\micro\meter} thickness and \vbias =\vtwo -\SI{30}{V}, the field amplitude increase by a factor two, resulting in a 50\% increase in the electron velocity. Such a modification would lead to $\mathrm{T_{e}}~\sim$ \SI{170}{\pico\second} thus $\Delta_{t}\sim$ \SI{15}{\pico\second}.\\
 The time resolution of the sensor will also be affected by the fact that the drift distance towards the induction region \induction will vary as a function of the lateral position of the primary charge carrier in \drift. Indeed, as the primary charge carriers will drift along the electron velocity streamlines visible in Figure~\ref{fig:fieldlines:efield}, their total drift distance will vary as a function of their initial transverse position. The in-homogeneity in the drift path depends on the ratio of the pitch to the pillar width. The lower it is, the smaller the in-homogeneity. The standard deviation of the drift path as function of the transverse position of the primary electrons is illustrated in Figure~\ref{fig:pitch_scan}. For the default \SI{2}{\micro \meter} width and \vbias =\vtwo -\SI{30}{V}, it varies from \SI{0.3}{\micro \meter} to \SI{1}{\micro\meter} for pitch varying from \SI{6}{\micro \meter} to \SI{15}{\micro\meter}. The standard deviation of the drifting time for uniform transverse position distribution then varies from \SI{5}{\pico \second} to \SI{23}{\pico\second}. \\
 This contribution is sub-leading with respect to the contribution from the centroid time standard deviation which may range from 15 to \SI{40}{\pico\second} depending on the thickness of \drift and the electric field amplitude in this region.

\section{Discussion and future work}
\label{sec:discussion}
\begin{figure}
        \centering
        \includegraphics[width=\textwidth]{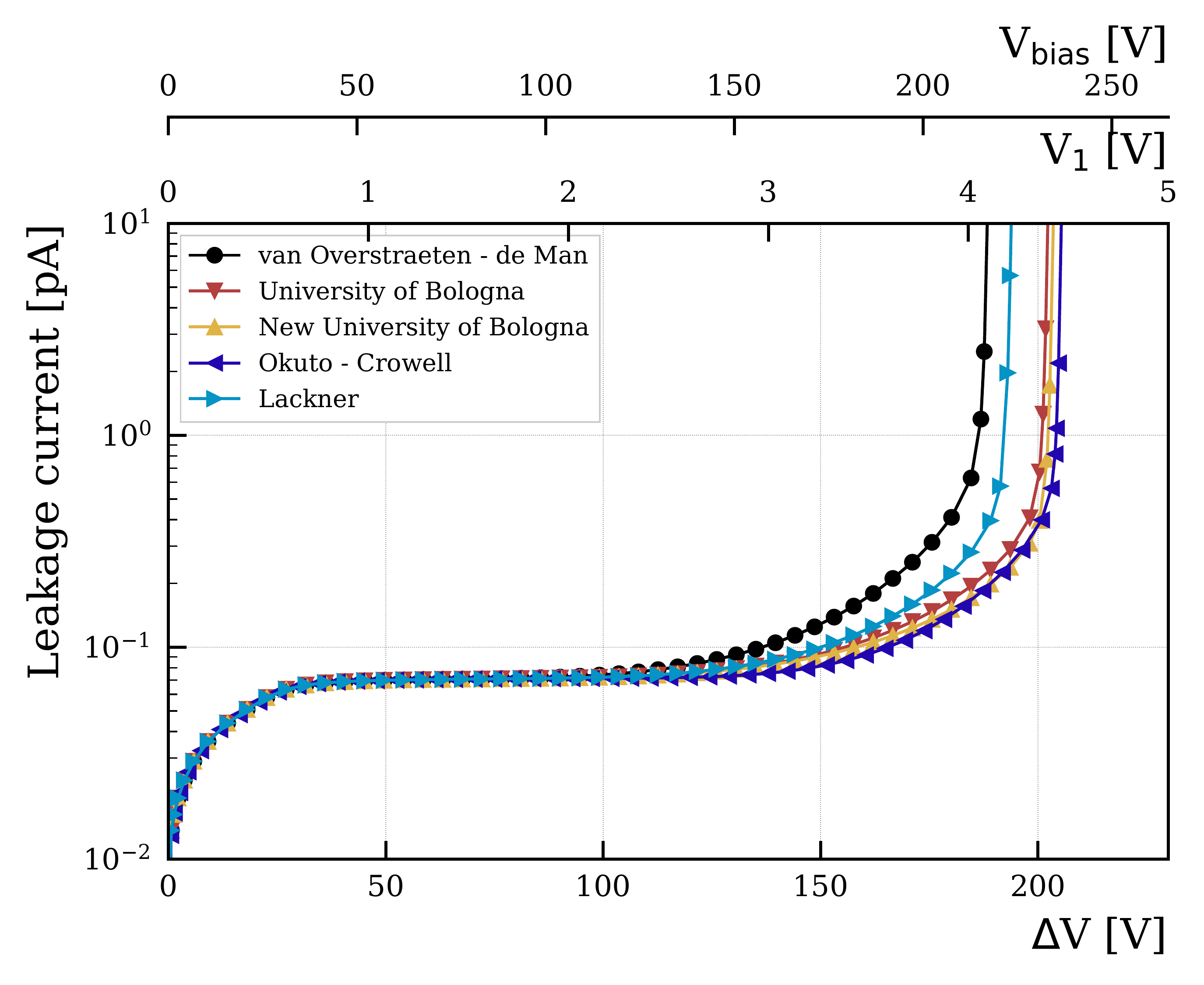}
        \caption{IV-curves of quasi-stationary simulations for different impact ionisation models.}
    \label{fig:impact_ionisation_models}
\end{figure}

\paragraph*{Impact ionisation model}
In the results presented so far, the van Overstreaten - de Man impact ionisation model~\cite{vanoverstreaten} has been used.
Several models are available and the I-V characteristic curve shown in Figure~\ref{fig:ivcurve} have been recalculated using the University of Bologna~\cite{unibo}, and the "New" University of Bologna~\cite{newunibo} models, as well as the Okuto - Crowell\cite{okuto} and the Lackner~\cite{lackner} models for impact ionisation. The results are compared in Figure~\ref{fig:impact_ionisation_models}. The I-V characteristics and all the curves show similar increase of the leakage current, corresponding to charge multiplication, over about a \SI{50}{\volt} range before reaching breakdown conditions. The main difference between the various models is the values of \dv at which charge multiplication appears, with the van Overstreaten - de Man model predicting charge multiplication about \SI{20}{\volt} earlier than the other models. 

\paragraph*{Response to minimum ionising particles}
A complete estimation of the time resolution that can be achieved by such sensor requires on one hand to simulate the interaction of MIP with the sensor, and on the other hand to study the impact of the front-end electronics on the signal. The MIP simulation is performed with \texttt{Garfield++}~\cite{garfield}. A full study of the impact of the sensor geometry and field configuration will be the topic of a future publication, but preliminary studies have been performed to understand the compatibility of the gain estimated in TCAD from a charge cloud deposited in the center of the bulk, as described in section~\ref{sec:simulation:amplification}, and the gain estimated from the amount of charge induced on the readout electrode by a MIP going through the detector with and without considering impact ionisation. In Figure~\ref{fig:gap_study}, the result of the gain evaluated with \texttt{Garfield++} is compared to the gain evaluated with the transient simulation of Sentaurus TCAD, using an identical setup, ie. the deposit of a charge cloud in the center of region \induction. For the configuration which have been studied, d=\SI{3}{\micro\meter} and w$_1$=\SI{3}{\micro\meter}, the gain estimation with the two methods agree to within five percent. 


\paragraph*{Number of electrodes}
In order to provide further amplification and a better control of the field in the pillar, and providing the guard distance between the electrode and the pillar wall can be respected, geometries can be considered with more than two stacked amplification electrodes. Conversely, simpler geometries with a single electrode are also interesting, as they could be made with simpler fabrication flow and no more overlay constraint between \electrodeone and \electrodetwo. As shown in the discussion on electrode geometry, namely on Figure~\ref{fig:retraction}, a single electrode geometry can indeed achieve gain. In those configurations, the smaller the guard between the electrode and the pillar wall, and the smaller the width over height pillar ratio, the better. It prevents highly peaked electric field on the edge of the pillar just above the readout electrode. 

\paragraph{Alternative materials} Any semi-conductor could be used for the bulk provided the band-gap is low enough to produce significant amount of charge carriers. Whereas depleted silicon offers more charges per MIP for low leakage current, silicon carbide and diamond are intrinsically more radiation resistant and while less electron holes pairs from ionisation by MIPs are expected, the amplification structure could compensate for it. 

\paragraph*{Alternative processes} The default process, based on DRIE, will be studied in the coming year. In parallel, an alternative process based on metal assisted chemical etching in liquid or gaseous phase\cite{metaletch,gasetch}, followed by plating of the etching catalyst will be studied. It could achieve high aspect ratio pillars without significant guard distance between the pillar wall and the electrode. This process could be ideal for single electrode geometries. 
In the case of diamond bulk, graphitisation of the diamond by laser could be used to produce the embedded electrode as is the case in 3D sensor applications~\cite{3Dgraphit}.

\paragraph{Possible adverse effects} It is possible that DRIE etching generates low quality silicon surface terminations, leading to higher leakage currents. Annealing of the surface and slow growth of a thermal oxide should allow this effect to be minimise. Another concern is that filling the etched area with silicon oxide could produce adverse effect after irradiation due to ionization in the oxide. To some extent the build up of charge at the silicon interface should be prevented by the presence of the buried amplification electrodes who may collect the slow moving holes in the oxide before they reach the interface. This will be further studied in simulation. Indeed, if the dielectric is strictly needed in between \electrodeone and \electrodetwo in the double electrode configuration, filling up the rest of the etched region is expected to ease the under bump metallisation deposition, but alternative techniques can be investigated,\\
The production of a demonstrator will be a key achievement to study the impact of those effects and assess the real performance of the device.
\section{Conclusion}
A new approach to achieve internal gain in solid state detector has been presented. The Silicon Electron Multiplier (SiEM) combines a conversion region where the passage of a MIP creates charge carriers, and an amplification and induction region where the charge multiplication occurs and the moving charges induce signal on the readout electrode. Simulations are used to assess and optimise the relevant geometry variables. The expected performance is estimated through simulations as well for a geometry that could be implemented in silicon with a process based on deep reactive ion etching.

\section*{Acknowledgment}
The authors would like to thank warmly Jakob Haimberger and Heinrich Schindler for their help with the simulations as well as Marc-Olivier Bettler for his careful proofreading This work was carried out in the context of the CERN Strategic R\&D Programme on Technologies for Future Experiment~\href{https://ep-rnd.web.cern.ch/}{https://ep-rnd.web.cern.ch/}.

\bibliographystyle{elsarticle-num} 
\bibliography{main}

\end{document}